\documentclass[sigconf]{acmart}

\usepackage{booktabs} 
\usepackage{url}

\setcopyright{acmcopyright}

\setcopyright{acmlicensed}
\acmConference[DAPA '19]{DAPA 2019 WSDM Workshop on Deep matching in Practical Applications}{February 15th, 2019}{Melbourne, Australia} 
\acmYear{2019}
\copyrightyear{2019}
\acmDOI{}
\acmISBN{}
\acmPrice{}
\begin{document}
\title{EENMF: An End-to-End Neural Matching Framework for E-Commerce Sponsored Search}

\author{Wenjin Wu, Guojun Liu, Hui Ye, 
	Chenshuang Zhang, Tianshu Wu, Daorui Xiao, Wei Lin, Xiaoyu Zhu}
\orcid{1234-5678-9012}
\affiliation{%
  \institution{Alibaba Group}
}
\email{{kevin.wwj,guojun.liugj,yehui.yh,chenshuang.zcs,shuke.wts,daorui.xdr,yangkun.lw,benjamin.zxy}@alibaba-inc.com }

\begin{abstract}
E-commerce sponsored search  contributes an important part of revenue for the e-commerce company. In consideration of effectiveness and efficiency, a large-scale sponsored search system commonly adopts a multi-stage architecture. We name these stages as \textit{ad retrieval}, \textit{ad pre-ranking} and \textit{ad ranking}. \textit{Ad retrieval} and \textit{ad pre-ranking}  are collectively referred to as \textit{ad matching} in this paper. In the \textit{ad matching} stage, there are two important problems that need to be addressed. First, in the keyword-based mechanism of traditional sponsored search, it is a great challenge for advertisers to identify and collect lots of relevant bid keywords for their ads. Due to the improper keyword bidding, advertisers cannot get their desired ad impressions; meanwhile, sometimes there are no ads displayed to user for long-tail queries. These issues lead to inefficiency. Second, deep models with personalized features have been successfully employed for click prediction in the ranking stage. However, for the reason of computing complexity, deep models with personalized features are not effectively and efficiently applied in the \textit{ad matching} stages. To address these two problems, we propose an end-to-end neural matching framework (EENMF) to  model two tasks---\textit{vector-based ad retrieval} and \textit{neural networks based ad pre-ranking}. Under the deep \textit{matching} framework,  \textit{vector-based ad retrieval} harnesses user recent behavior sequence to retrieve relevant ad candidates without the constraint of keyword bidding. Simultaneously, the deep model is employed to perform the global pre-ranking of ad candidates from multiple retrieval paths effectively and efficiently. Besides, the proposed model tries to optimize the pointwise cross-entropy loss which is consistent with the objective of predict models in the ranking stage. We conduct extensive evaluation to validate the performance of the proposed framework. In the real traffic of a large-scale e-commerce sponsored search, the proposed approach significantly outperforms the baseline.

\end{abstract}

\keywords{Sponsored search, Ad retrieval, Ad matching, Ad pre-ranking, Deep learning}
\maketitle
\section{Introduction}
When users search in the search engine, sponsored search platform enables advertisers to target advertisements (ads\footnote{In the remainder, ad(s) is used to refer to advertisement(s)}) to users' search requests. Along with organic results, search engine presents the sponsored results to users in response to their search requests. Precisely, in e-commerce sponsored search, organic results are the products named \textit{"item"} on Taobao platform, while the ads are also a special kind of \textit{item}.


In consideration of effectiveness and efficiency, large-scale search systems or recommendation systems often adopt a multi-stage search architecture~\cite{Liu:2017:CRO:3097983.3098011}. In our e-commerce sponsored search system, a three-stage architecture has been adopted over the past few years. Sequentially we name these three stages as \textit{ad retrieval}, \textit{ad pre-ranking} and \textit{ad ranking}.
In this paper, we refer the ad \textit{ retrieval}  and \textit{pre-ranking}  stage both as \textit{matching}. We mainly focus on proposing an efficient and effective neural matching model for the ad \textit{matching} stage in sponsored search.

In our e-commerce sponsored search, there exist two types of problems in the \textit{ad matching} stage. Firstly, in the traditional sponsored search system, keyword-based mechanism provides a simple ad retrieval method where the whole burden is on advertisers, making it a big challenge for advertisers to optimize bids. It is quite impossible for advertisers to identify and collect lots of relevant bid keywords to target their ads. Due to the improper keyword bidding, advertisers possibly can not get the desired ad impressions; meanwhile, there are no ads displayed in the search result pages for long-tail queries. To alleviate this problem, the search engines often provide an advanced matching service to advertisers, which rewrites user's original query to many different related bid keywords, enriching the connections between user's query and bid keywords of ads. The query rewriting approach is limited to matching only against predefined bid keywords of ads. Thus, the keyword-based mechanism is still unable to achieve a good match between user query requests and advertisements. Secondly, various types of personalized information such as user profile, user long-time click behaviors and real-time click behaviors, have been proved to be effective for click prediction models in the ranking stage~\cite{conf/wsdm/ChengC10,conf/rnnctr,conf/deepcrossing}. However, for the sake of computing complexity, deep models exploiting personalized information such as user recent click behaviors, is not utilized in the \textit{matching} stage. 

To address the inefficiency of keyword-based mechanism and lack of uniformly user personalized information modeling in the \textit{matching} stage, inspired by recent work on multi-task learning ~\cite{conf/naacl/LiuGHDDW15,Bai-MTL}, we propose a practical neural \textit{matching} model to fulfill these two tasks: \textit{vector-based ad retrieval} and \textit{neural networks based ads pre-ranking} respectively. \textit{Vector-based ad retrieval} exploits user recent behavior sequence to select relevant ad candidates without the constraint of keyword bidding. For the vector retrieved ads without bidding information, a bid optimizing strategy called Optimized Cost Per Click (OCPC)\cite{conf/www2017/Yansu} is applied to determine how much the advertisers will be charged if their ads are clicked.  Simultaneously, there often exist various ads retrieval paths in sponsored search system such as keyword-based retrieval and vector-based retrieval. We adapt the model to perform the global pre-ranking of ad candidates. The model is trained to optimize the cross-entropy loss under the guide of search impression logs, which makes the optimized objects of matching and ranking stages consistent. Finally, we conduct online evaluation in the real-world operational environment of our large-scale e-commerce sponsored search. 
 
The main contributions of this paper are summarized as follows:
\begin{itemize}
	\item We propose a novel multi-task neural matching framework for a large-scale e-commerce sponsored search, which is trained on user search sessions. The proposed matching model tries to optimize the cross-entropy loss function which is consistent with the objective of predicting models in the ranking stage.
	
	\item Under the framework, we implement the \textit{vector-based ad retrieval} to overcome the shortness of the keyword-based ad retrieval and provide advertisers a keyword-free way to advertise their products.
	
	\item The proposed approach is deployed in a large-scale e-commerce sponsored search platform. The online evaluation reveals that the proposed method retrievals and selects more relevant ads than the baseline methods.

\end{itemize}


\begin{figure}
	\includegraphics[width=0.5\textwidth]{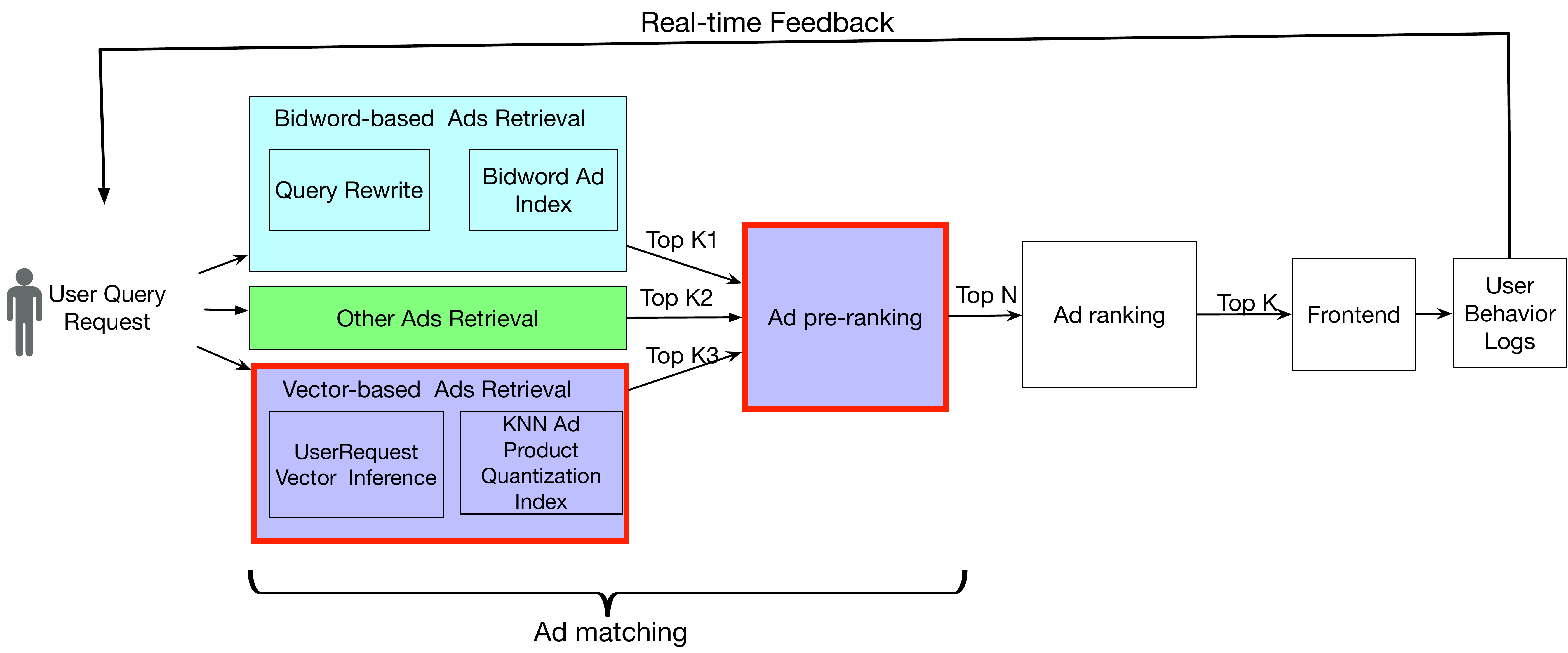}
	\caption{An E-commerce Sponsored Search System Overview. This paper focuses on two parts: \textit{vector-based ad retrieval} and \textit{ad pre-ranking}, which are highlighted in red. \textit{Ad retrieval} and \textit{ad pre-ranking}  are collectively referred to as \textit{ ad matching} in this paper.}
	\label{fig:architecture}
\end{figure}

\section{Methodology}\label{section:method}

\begin{figure}
	\includegraphics[width=0.5\textwidth]{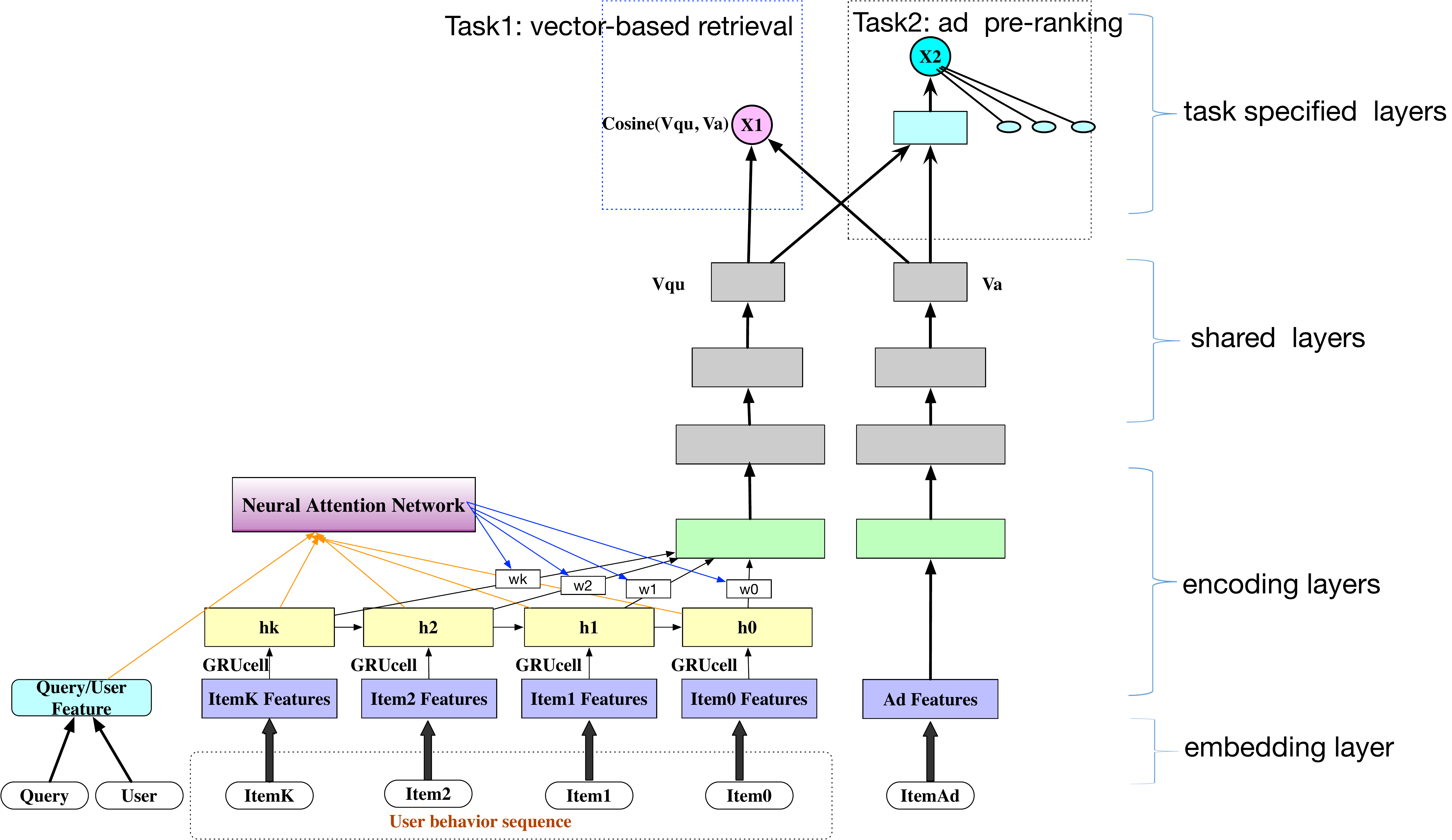}
	\caption{Neural Matching Model. Attentive GRU-RNN is adapted to model user behavior sequence, and this model fulfills two tasks: \textit{vector-based retrieval} and \textit{deep ad pre-ranking}. }
	\label{fig:model}
\end{figure}

 In order to better understand the proposed neural matching framework, Figure \ref{fig:architecture} the overall architecture and data flow of the sponsored search system. 

\subsection{Model Architecture}
The proposed model architecture is shown in Figure \ref{fig:model}. Horizontally, the architecture consists of two parallel sub neural networks, one network for user requests (which we term \textit{Qu-Net}) in the left side and the other for advertisements (which we term \textit{Ad-Net}) in the right side. User request features and advertisement features are fed to \textit{Qu-Net} and \textit{Ad-Net} as inputs respectively, and ad clicked labeled 1 or not clicked labeled 0 is produced as output. Vertically, the underlying model architecture can be divided into four parts from bottom to top including input and embedding layers, encoding layers, shared layers and task-specified layers. We detail these layers in the following.



\subsection{Input and embedding layers}
 The input instance of the proposed models is consisting of four types of features: query feature, user profile feature, user previous behaviors feature and the target ad item feature. User behaviors is a behavior sequence $ X=\left\lbrace q_{0},i_{1},i_{2},...,i_{m} \right\rbrace $ where $ q_{0} $ is user's current query, and $i_{k}$ indicates the $k$th behavior item that the user acted before this search request. $X$ is ordered by the time of user behavior. Each behavior item is represented by ID features including item ID, shop ID, brand ID, term IDs of the item's title and the corresponding search query feature. In \textit{Ad-Net}, ad item is also represented by ID features like the behavior item, except the search query feature. Each unique ID space has a separately learned embedding matrix. Very large cardinality ID space (e.g. product item Ids and search query terms) are truncated by keeping top ones after sorting based on their frequency in search logs. Out-of-vocabulary values are set to zero and mapped to the embedding of zero. Multivalued ID feature embeddings, such as word IDs of item title, are summed before fed to the next layer. 

Remarkably, sparse features in the same ID space share the same underlying embedding matrix. Since an ad item is also a product, ad item features in  \textit{Ad-Net} share all embedding matrices with behavior item features in \textit{Qu-Net}. Sharing embedding is important for improving generalization, speeding up model training and reducing model parameters. 

\subsection{Encoding layers}
When a user search in e-commerce mobile app, she browses and clicks product items in the form of streaming. For instance, when a user want to buy a shoes and search "shoes" , she usually browses the result list, clicks shoes which she likes and compares them before adding to cart. Intuitively, items in the same behavior sequence are correlated. In other words, user previous behavior items are predictable for the next behavior item. This type of relation has been proved to be effective in recommendation system ~\cite{conf/recsys/YoutubeCovingtonAS16,deepinterest}

In the encoding layer, we use recurrent neural network (RNN) to encode user behavior sequence. We consider the latest previous $m$ behaviors, padding the default symbol to the fixed size m if length of previous behaviors is less than $m$.  We adopt GRUs, since GRUs have fewer parameters and competitive performance to LSTMs\cite{GRU2ChungGCB14}.

In the e-commerce sponsored search, there may exist items in previous behavior sequence unrelated to current search query. For example, user current search query is "red dress", while in her previous behavior sequence there are dress product items searched by "dress" and shoes product items searched by "shoes". Obviously, these two category of product items are of different relevance to current search query "red dress". Thus, we adopt query based attention nets to address this problem. Vector $h_{j}$ is the GRU hidden output at the step $j$. We take $h_{j}$ as the representation of $j$-th behavior item, and represent the behavior sequence as a weighted sum of the vector representation of all behavior items. The attention weight makes it possible to assign proper credit to items according to their importance to current query request. Mathematically, it takes the formulates as follows:

\begin{equation}
	h=\sum_{t=1}^{m} w_{t}h_{t}
\end{equation}

\begin{equation}
	w_{t}=\dfrac{\exp (net(h_{t},Q;\theta))}{\sum_{i=1}^{m}\exp (net(h_{i},Q;\theta))}
\end{equation}

where $w_{t}$ is the weight for $h_t$, $net(h_{t},Q;\theta)$ is a two-layer attention network parameterized by $\theta$ with hidden state $ h_{t} $ and query feature embeddings $ Q $ as inputs. $ h $ is the vector representation of user previous behavior items.
As for \textit{Ad-Net}, we directly use one fully connected layer to map ad item embeddings to a vector with the same dimension as the encoding vector of user previous behavior items $ h $.

\subsection{Shared layers}\label{subsection:sharedlayers}
After the encoding layers, we get a user query request vector output and an ad vector output with the same dimension, which may not well fit in the same vector space. Inspired by DSSM models~\cite{conf/cikm/HuangHGDAH13_dssm}, we stack two shared nonlinear fully connected layers over the encoding layer of \textit{Qu-Net} and \textit{Ad-Net} to bind these two types of vectors. Besides, hard parameter sharing greatly reduces the risk of overfitting~\cite{journals/corr/Ruder17a}. Furthermore, let $ h_{l}$ denotes the corresponding output of the $ l $-th hidden layers. 
\begin{equation}
h_{l}=f(h_{l-1})
\end{equation} 
where $f(\cdot) $ is an non-linear activation function.
The output of the shared layers are query request's representation and ad's with dimension $d$ (e.g. 128), which are fed to the next multi-task specific layers. 

\subsection{Multi-task specific layers}\label{subsection:multi-task}
Through the previous layers, we obtain representations of both query request and ad. Our model has two tasks to fulfill: vector-based ads retrieval and ads pre-ranking. For these two different tasks, we apply specific network layers and optimization objectives.

\subsubsection*{\textbf{Vector-based ad retrieval}} For the \textit{vector-based ad retrieval} task, with $ V_{qu}$  and  $V_{a} $ as inputs the relevance score is computed by cosine similarity as:
\begin{equation}
 	cosine(V_{qu}, V_{a})=\dfrac{V_{qu} \cdot V_{a}}{||V_{qu}|| \cdot ||V_{a}||} 
\end{equation}
The larger the cosine value, the more similar is between $ V_{qu}$  and  $V_{a} $. We use the cross-entropy loss as the objective to train model:

\begin{equation}
	C_{v}(\theta)=-\frac{1}{N} \sum_{i=1}^{N}(y_{i}\log(P(V_{qu,i}, V_{a,i}))) + (1-y_{i}) \log(1-P(V_{qu,i}, V_{a,i})) 
\end{equation}

\begin{equation}
	P(V_{qu,i}, V_{a,i})=\dfrac{\exp(\gamma cosine(V_{qu,i}, V_{a,i}))}{1+\exp(\gamma cosine(V_{qu,i}, V_{a,i})}
	\label{eq:gamma}
\end{equation}
where $\gamma$ is a tuning factor determined by validation set. $ y_{i}\in\{0,1\} $ is the target. If user clicked the current ad, the instance is positive, otherwise negative.  The loss is 
summed over all samples in a mini-batch (128 samples in our experiments). At serving time, ad retrieval is reduced to a nearest neighbor search problem. Product quantization is used to implement K nearest neighbor search, which is efficiently supported in Faiss library~\cite{JDH17}. The details is that we apply forward inference of the current query request through \textit{Qu-Net} obtaining the $V_{qu}$ vector which is normalized, and then we use $V_{qu}$ to search ads' Faiss index to obtain relevant ads.

\subsubsection*{\textbf{Ad pre-ranking}} In our scenario, thousands of ads are recalled through bidword-based ad retrieval and vector-based retrieval. The ads pre-ranking stage needs to score and select top N (e.g. 200) ad candidates for the ranking stage. Different from the baseline approach which uses static score built in ads inverted index to select top N ads, we employ a deep model with personalized features to score and select top N ads. Here we still model it as a click-through rate prediction (CTR) problem, and use the cross-entropy loss as the objective which is consistent with the objective of CTR predicting model in the ranking stage. In order to model the interaction between query request features and ad features, we add a nonlinear fully connected layer for them. The loss function is descirbed in Equation \ref{eq:loss2}.
\begin{equation}
	C_{r}(\theta)=-\frac{1}{N} \sum_{i=1}^{N}(y_{i}\log(P'(V_{qu,i}, V_{a,i}))) + (1-y_{i}) \log(1-P'(V_{qu,i}, V_{a,i}))
	\label{eq:loss2}
\end{equation}

\begin{equation}
	 P'(V_{qu}, V_{a}) = FC(V_{qu}, V_{a};\theta)
\end{equation}
where the lightweight networks $ FC(V_{qu}, V_{a};\theta) $  is consisting of one fully connected layer and a logits regression layer, with the concatenation of $ V_{qu}$  and  $V_{a}$ as inputs. The reason for choosing $ FC(V_{qu}, V_{a};\theta) $ is concluded as follows. First, most of the matrix computation $(V_{qu},V_{a})W$ in the fully connected layer can be calculated offline, since as described in Equation \ref{eq:qu}, $ W $ is the parameters matrix of this layer, $(V_{qu}, \vec{\textbf{0}})W$ is computed only one time per query request and $(\vec{\textbf{0}}, V_{a})W$ can be computed offline for ads in advance. Second, the lightweight networks $ FC(V_{qu}, V_{a};\theta) $ are flexible to incorporate other effective features such as id features or statistic features, which is similar to the wide part of  Heng{-}Tze Cheng et al.'s deep \& wide model ~\cite{conf/widedeep}. 

\begin{equation}
 (V_{qu},V_{a})W = (V_{qu}, \vec{\textbf{0}})W + (\vec{\textbf{0}}, V_{a})W  
 \label{eq:qu}
\end{equation}

Finally, our model is trained jointly with the objective:

\begin{equation}
 	C_{joint}(\theta) = \alpha C_{v}(\theta) + (1-\alpha)C_{r}(\theta)
 \label{eq:jointloss}
\end{equation}
where $\alpha$ is a hyperparameter that balances the effects of two tasks.

\section{Experiments}\label{section:experiment}

\subsection{Dataset description}
We use the search logs from both sponsored search and organic search to reorder each user's historical behaviors according to the timeline, and then construct the train dataset and test dataset. An instance records the complete information about an ad impression including user profile, query, user recent behaviors and corresponding behavior of the current ad (click or non-click). And the window size of user recent behaviors is 6. In our experiment, if user clicked the current ad then the instance is positive, otherwise negative. About $3\times10^{9}$ instances from search logs per day are sampled as data set. We use samples of every three consecutive days for training and test on the samples of the next day. We divide the search logs of 12 consecutive days to three groups of training and testing data sets for training and evaluation. We employ distributed \textit{Tensorflow} \footnote{https://www.tensorflow.org/}machine learning platform deployed on a large-scale computing cluster to train our neural networks 

\subsection{Offline Evaluation}

\subsubsection{\textbf{Comparison of different network architectures}}
In order to investigate the effectiveness of the proposed model, we compare  five network architectures and the baseline model DSSM~\cite{conf/cikm/HuangHGDAH13_dssm}. These models are described as follows:  
\begin{itemize}
	\item \textbf{DNN}: it employs the mean pooling to represent user previous behaviors, ignoring the order information and taking all behavior items equally. 
	\item \textbf{GRU-RNN}: it applies GRU cell based RNN to model user previous behavior sequence.
	\item \textbf{Attention-DNN}: it adds a query based attention network over \textbf{DNN}, and distinguishes the different role that each behavior item plays when predicting the current interest.
	\item \textbf{Attention-GRU-RNN}: Similarily, it adds the query based attentive network \textbf{GRU-RNN}, considering both order and importance information.
	\item \textbf{Concatenate-DNN}: it directly concatenates embeddings of user previous behaviors, letting the raw information feed into later layers.
	\item \textbf{DSSM}~\cite{conf/cikm/HuangHGDAH13_dssm}: In DSSM~\cite{conf/cikm/HuangHGDAH13_dssm}, a query is parallel to the titles of the ad documents clicked on for that query. We extracted
	the query-title pairs as positive samples for model training from ads click logs using a procedure similar to ~\cite{conf/cikm/HuangHGDAH13_dssm}, and randomly sampled four negative documents per positive sample. Since user's search query is often short, we enrich the query with the title of recent behavior items. We also use terms in the query and ad's title as input features. 
	\item \textbf{Search2Vec} ~\cite{grbovic2016scalable}, which is the state-of-art approach for board match in sponsored search. Since our offline evaluation is based on the search session logs and user request is sparse, we do not conduct offline evaluation for Search2Vec~\cite{grbovic2016scalable}. However, we trained the Search2Vec model and conducted the online evaluation with the real search traffic, which is described in Subsection \ref{subsection:online_evaluation}
\end{itemize}

To measure the overall performance of each model, we employ Area Under ROC Curve (AUC) as the evaluation metric, which is widely used in industry. AUC measures whether the clicked instances are ranked higher than the non-clicked ones. For the fair model comparison, we tune model parameters using validate dataset (5\% samples from training dataset not used for training models) to ensure these models to achieve its best performance respectively. 

\begin{table}
	\caption{Comparison of Different Models for User Behavior Sequence on Task1 \textit{vector-based ads retrieval} and Task2 \textit{ads pre-ranking}}
	\label{tab:comparison_models}
	\begin{tabular}{lcc}
		\toprule
		model type & Task1 AUC  & Task2 AUC\\
		\midrule
		 DNN & 0.6657 & 0.6655 \\
		 GRU-RNN & 0.6760 & 0.6758 \\
		 Attention-DNN & 0.6762 & 0.6760 \\
		 Attention-GRU-RNN & \textbf{0.6885} & \textbf{0.6886} \\
		 Concatenate-DNN & 0.6795 & 0.6796 \\
		 DSSM~\cite{conf/cikm/HuangHGDAH13_dssm} & 0.6200 & - \\
		\bottomrule 
	\end{tabular}
\end{table}

Table ~\ref{tab:comparison_models} reports the overall AUC of all models on the test dataset. These results demonstrate that RNN based models outperform DNN based models, and models with attention mechanism outperform the ones without respectively for both tasks. Specifically, RNN brings about 0.01 AUC improvement comparing with DNN. Attention mechanism brings about 0.01 AUC improvement for DNN and 0.012 for RNN. Concatenate-DNN outperforms DNN with about 0.01 AUC improvement. These results conform the hypothesis that user previous behavior items sequence are predictable for the next behavior item, but previous behavior items are not equally important. The overall evaluation results show the effectiveness of the GRU-RNN with attention model. Besides, the result of the original DSSM~\cite{conf/cikm/HuangHGDAH13_dssm} method indicates that it is not fit for ad retrieval very well. There are two reasons for that. First, DSSM~\cite{conf/cikm/HuangHGDAH13_dssm} solely employing the term features can not distinguish between positive and negative samples very well, and we find that in the training dataset the positive and negative samples seem to be relevant to their corresponding query in the textual content. Second, the DSSM~\cite{conf/cikm/HuangHGDAH13_dssm} method employs a negative sampling based pairwise loss which does not directly aim at optimizing CTR prediction.

In the following, we conduct more detailed analysis of our model in order to analyze the individual effect of different components or parameters on the performance. In each experiment, we only check one component or parameter, while the rest will be fixed.

\subsubsection{\textbf{Influence of jointly training}}
We compare our proposed jointly training model with the single training model. As described in Section \ref{subsection:multi-task}, our model is trained to fulfill two tasks: \textit{vector-based ads retrieval} and \textit{ads pre-ranking}. We conduct this comparison based on the GRU-RNN with attention model. As shown in Table \ref{tab:multitask}, the jointly training model achieves better performance than the single trained ones. For the \textit{vector-based ads retrieval} task, the joint model leads to about 0.0120 AUC improvement. As for the ads pre-ranking task, the joint model leads to about 0.0088 AUC improvement. Besides, when these two tasks are trained individually, \textit{Task2} 's AUC is larger than \textit{Task1}'s, which is consistent with the empirical idea that a lightweight networks $ FC(V_{qu}, V_{a};\theta) $ is more powerful than the cosine interaction between $ V_{qu} $ and $ V_{a} $. One possible reason is that the joint training tends to learn more expressive representations of user request and ads. However, we find that two tasks almost have no significant difference in AUC metric when they are jointly trained. 

\begin{table}
	\caption{Comparison of jointly traning and single training on two tasks: Task1 \textit{vector-based ads retrieval} and Task2 \textit{ads pre-ranking} }
	\label{tab:multitask}
	\begin{tabular}{lcc}
		\toprule
		& Task1 AUC  & Task2 AUC\\
		\midrule
		single training task1 & 0.6765 & - \\
		single training task2 & - & 0.6798 \\
		jointly training & \textbf{0.6885} & \textbf{0.6886} \\
		\bottomrule 
	\end{tabular}
\end{table}
Importantly, when the sponsored search system serves online, these two tasks (\textit{vector-based ads retrieval} and \textit{ads pre-ranking}) are needed to serve together. Consideration of online serving's efficiency and convenience, one joint model is a better choose than two single models.

\subsubsection{\textbf{Influence of shared layers}}
In order to analysis the effect of shared layers described in Section \ref{subsection:sharedlayers}, we carry out the comparison experiment and the result is shown in Table ~\ref{tab:sharelayer}. It can be observed that share layers are consistently better than the non-share ones in both tasks. In one sense, sharing layers is a type of interaction between query request and ads. Besides, hard parameter sharing greatly reduces the risk of overfitting ~\cite{journals/corr/Ruder17a}.

\begin{table}
	\caption{Comparison of share vs. non-share layers for Task1 \textit{vector-based ads retrieval} and Task2 \textit{ads pre-ranking}}
	\label{tab:sharelayer}
	\begin{tabular}{lcc}
		\toprule
		& Task1 AUC  & Task2 AUC\\
		\midrule
		share & 0.6820 & 0.6821 \\
		non-share & 0.6761 & 0.6794 \\
		\bottomrule 
	\end{tabular}
\end{table}

\subsubsection{\textbf{Influence of hyperparameter $\gamma$}}
For the vector-based ad retrieval task, the relevance score between $V_{qu}$ and $ V_{a} $ is computed by cosine similarity. 
As shown in Equation \ref{eq:gamma} of Section 4.5, $P(V_{qu,i}, V_{a,i})$ is the predict value. 
We train the models using different $\gamma$ values and evaluation results are shown in ~\ref{tab:lambda}. The performance of model is very sensitive to parameter  $\gamma$. When parameter  $\gamma$ is 6, the model achieves the best performance. The mean of the final predict value is 0.0570 and minimum and maximum values are 0.0060 and 0.3000 respectively. And we find that the mean predict value 0.0570 is nearest to the ratio of positive instances in the training dataset.

\begin{table}
	\caption{Comparison of different parameter  $\gamma$ for \textit{vector-based ad retrieval task}}
	\label{tab:lambda}
	\begin{tabular}{lcl}
		\toprule
		 $\gamma$ & (mean,var,[min,max]) of predict value  &  AUC\\
		\midrule
		1 & (0.2690, 0.0000, [0.2690, 0.2710]) & 0.4999 \\
		3 & (0.0680, 0.0260, [0.0475, 0.2300]) & 0.6642 \\
		6 & \textbf{(0.0570, 0.0443, [0.0060, 0.3000])} & \textbf{0.6866} \\
		9 & (0.0580, 0.0300, [0.0120, 0.2400]) & 0.6533 \\
		\bottomrule 
	\end{tabular}
\end{table}

\subsection{Online Evaluation}\label{subsection:online_evaluation}
In this subsection we conduct online A/B testing on the e-commerce sponsored search platform with 1\% of overall search traffic lasting three days. Four metrics are used to evaluate the performance of the proposed approach as following: 

\begin{itemize}
	\item $CTR=AdClickCount/AdPresentCount $
	\item $PR=AdPresentCount/RequestCount$
	\item $CPC=AdCostAmount/AdClickCount$
	\item $RPM=CTR*PPC$
\end{itemize}
We deploy the proposed model EENMF for these two tasks: \textit{vector-based ads retrieval} and  \textit{ads pre-ranking} in the system.  

\subsubsection{\textbf{Evaluation of \textit{vector-based ads retrieval}}}
In our system, there exists two retrieval methods: the first one is a graph covering based query rewriting method similar to \cite{nb} used to retrieval the keyword-based ads; and the second one is BKR  which is a variant of the method proposed in ~\cite{conf/www2017/Yansu} for broad match. Search2Vec ~\cite{grbovic2016scalable}, which is the state-of-art approach for board match in sponsored search. Therefore, we conduct four groups of experiments: (1) Search2Vec; (2) BKR; (3) EENMF; and (4) EENMF combines with BKR. To compare fairly, all these methods employ the same user behavior information including queries and recent clicked items to recall ads. In the online A/B test, users, together with their queries, are randomly and evenly distributed to four buckets. Each experimental group is deployed in one bucket.

As shown in Table ~\ref{tab:online1},
The improvement of metrics PR and CTR demonstrates that all these keyword-free and vector-based ad retrieval methods can recall more and better ad candidates in the ad matching stage. The lifts also illustrates that advertisers are able to receive more valuable traffic even if they choose the keyword-free bidding advertising. Meanwhile, the lift of RPM metric indicates the improvement of the sponsored search platform's monetization ability. Specifically, EENMF outperforms Search2Vec significantly. ENMF's performance is a little better than BKR. The reason may be that EENMF model is deeper and more expressive than both Search2Vec and BKR, and the optimized objective of EENMF is more consistent with the object of ranking stage. Besides, the combination of EENMF and BKR achieves higher PR and RPM metrics, which is valuable to the ad platform.

\subsubsection{\textbf{Evaluation of \textit{ads pre-ranking}}}
As for the \textit{ads pre-ranking} task, the baseline is a heuristics method based on the static ad-level quality scores from indexes and Jaccard similarity between query and ad in categories and properties.

\begin{table}
	\caption{Comparison of Average Online Metric Lift Rates for Task1 \textit{vector-based ads retrieval}}
	\label{tab:online1}
	\begin{tabular}{lcccc}
		\toprule
		Methods &  PR & CTR & CPC & RPM \\
		\midrule
		Search2Vec~\cite{grbovic2016scalable} & 2.2\%  & 0.5\%  & 1.9\% & 2.4\% \\
		BKR ~\cite{conf/www2017/Yansu}        & 2.1\%  & 2.0\%  & 2.1\% & 4.1\% \\
		EENMF        & 2.5\%  & 1.9\%  & 3.2\% & 5.1\% \\
		EENMF + BKR ~\cite{conf/www2017/Yansu}  & 3.5\%  & 2.7\%  & 4.0\% & 6.7\% \\
		\bottomrule 
	\end{tabular}
\end{table}

As shown in Table \ref{tab:online}, the CTR increases 3.1\% and  CPC decrease 3.3\%, which proves the effectiveness of the proposed pre-ranking deep model. Overall, these online evaluation results demonstrate the significant effectiveness of the proposed approach.

\begin{table}
	\caption{Average Online Metric Lift Rates for  Task2 \textit{ads pre-ranking}}
	\label{tab:online}
	\begin{tabular}{lc}
		\toprule
		Metric &  Lift Rate\\
		\midrule
		PR  & 0.0\% \\
		CTR & 3.1\% \\
		CPC & -3.3\% \\
		RPM & -0.2\% \\
		\bottomrule 
	\end{tabular}
\end{table}

\section{Discussion} \label{section:discuss}
In this section, we discuss the efficiency of online inference. In a large-scale e-commerce sponsored search system, it is critical to response to user query request timely. Usually, given a query request, it is very time-consuming to perform ad ranking in a deep learning based sponsored search system. Since there exist quite a lot of ad candidates in the matching stage, the proposed deep models need to be served online efficiently. Reviewing the model architecture described in Section \ref{section:method}, we  find that it is divided to two sub networks: \textit{Qu-Net} and \textit{Ad-Net}. When the model serves online request, for the user query request side, we just need one forward inference per query request through \textit{Qu-Net}  and obtain the $V_{qu}$ vector. For the ad side, we conduct forward inference for all ads in the advertising repository using the trained network model offline and these inferenced ad vectors $V_{a}$ are built into index. When a new ad arrives, its vector $V_{a}$ is infereced and written into index in the updating system. 


\section{Related work}\label{section:relatedwork}

Query rewriting has been well studied in the literature. These works on query rewriting fall into two categories: one is based on the relevance matching among queries and ads~\cite{conf/cikm/BroderCFGJR08, conf/rarequeries}, and the other is based on mining the co-occurrence among queries and ads from the historical ad click logs ~\cite{journals/corr/Sussillo14, conf/query-flow-graph}. However, these methods can not overcome the limitation of keyword-based ad retrieval.  To address these problems, the most related works were conducted by Mihajlo Grbovic \textit{et al.}  ~\cite{grbovic2016scalable} . Mihajlo Grbovic \textit{et al.}~\cite{grbovic2016scalable} designed a query-ad semantic matching approach based on embeddings of queries and ads, namely Search2Vec. The embeddings were learned by the skip-gram model~\cite{conf/mikolov2013distributed} on user search session data in an unsupervised manner. Different from their works, we propose a deep leaning based matching framework to realize the vector-based ad retrieval and the global ad pre-ranking in an end-to-end manner, which is more compatible with new and long-tail ads and queries.

Another related work is  leveraging deep learning techniques for the semantic matching problem in the information retrieval and recommendation systems. In the context of information retrieval, many representation focused approaches based on the Siamese architecture
have been explored especially for short text matching, such as DSSM ~\cite{conf/cikm/HuangHGDAH13_dssm}. DeepCrossing model ~\cite{conf/deepcrossing} are some of the methods that learn query and document text embedding to predict click-through rate. Zhang \textit{et al.}~\cite{conf/rnnctr} proposed a framework directly modeling the dependency on user's sequential behaviors into the click prediction process through the recurrent network. The interaction focused neural models  ~\cite{Guo:2016} learn query-document matching patterns from word-level interactions. Wasi Ahmad \textit{et al.} ~\cite{ACW18} proposed a joint 
framework trained on search sessions to predict next query and rank corresponding documents. Our work falls into the representation focused approach to learn user request and document representations and jointly models the two tasks: retrieval ads and pre-ranking ads in sponsored search. 
In the context of recommendation, deep learning based models learn a better representation of user's demands, item's characteristics and historical interactions between them. Cheng \textit{et al.}~\cite{conf/widedeep}  proposed an App recommender system for Google Play with a wide \& deep model. Covington \textit{et al.}~\cite{conf/recsys/YoutubeCovingtonAS16} posed recommendation as extreme multiclass classification and  presented a deep neural network based recommendation algorithm for video recommendation on YouTube. DeepIntent model proposed in Zhai \textit{et al.} ~\cite{Zhai:2016:DLA:2939672.2939759} comprises a Bidirectional
Recurrent Neural Network (BRNN) combined with an attention module. Among these previous works, DeepIntent model~\cite{Zhai:2016:DLA:2939672.2939759} and DSSM~\cite{conf/cikm/HuangHGDAH13_dssm} are mostly similar to our work. However, There are two important different aspects. First, our model tries to minimize the pointwise cross-entropy loss which is consistent with the objective of CTR predicting model in the ranking stage. Second, we focus on employing an attention based GRU-RNN model to learn user search request which consists of user query and recent behavior sequence.

\section{Conclusions and Future Work} \label{section:conclusions}
The paper contributes an efficient and effective ad matching framework based on neural networks for the ad matching phrase in large-scale e-commerce sponsored search. The optimized objective of the proposed matching model is consistent with the predict model in the ranking phrase, which makes the performance of the multi-stage architecture better.  The neural network model introduces personalized and real-time features to the ad \textit{matching} stage. And we jointly fulfill the \textit{vector-based ad retrieval} task and the global \textit{ad pre-ranking} task in e-commerce sponsored search.  Comparing with baseline methods, experiment results show that our ad matching framework achieves better performance. In the near future, we will introduce more features to the model such as image features and explore external memory networks~\cite{memorynetwork} to model user behaviors.

\section*{Acknowledgment}
The authors would like to thank the colleagues for their valuable supports, such as Yan Zhang, Genbao Chen,  Yuping Jiang,  Hao Wan, Sheng Xu, Zhenkui Huang, Qing Ye, Tao Ma, Hang Xiang, Di Zhang, Hongbin Zhao, Jinhui Li, Bo Wu.


\bibliographystyle{ACM-Reference-Format}
\bibliography{sample-bibliography} 


\begin{thebibliography}{39}


\ifx \showCODEN    \undefined \def \showCODEN     #1{\unskip}     \fi
\ifx \showDOI      \undefined \def \showDOI       #1{#1}\fi
\ifx \showISBNx    \undefined \def \showISBNx     #1{\unskip}     \fi
\ifx \showISBNxiii \undefined \def \showISBNxiii  #1{\unskip}     \fi
\ifx \showISSN     \undefined \def \showISSN      #1{\unskip}     \fi
\ifx \showLCCN     \undefined \def \showLCCN      #1{\unskip}     \fi
\ifx \shownote     \undefined \def \shownote      #1{#1}          \fi
\ifx \showarticletitle \undefined \def \showarticletitle #1{#1}   \fi
\ifx \showURL      \undefined \def \showURL       {\relax}        \fi
\providecommand\bibfield[2]{#2}
\providecommand\bibinfo[2]{#2}
\providecommand\natexlab[1]{#1}
\providecommand\showeprint[2][]{arXiv:#2}

\bibitem[\protect\citeauthoryear{Ahmad, Chang, and Wang}{Ahmad
  et~al\mbox{.}}{2018}]%
        {ACW18}
\bibfield{author}{\bibinfo{person}{Wasi Ahmad}, \bibinfo{person}{Kai-Wei
  Chang}, {and} \bibinfo{person}{Hongning Wang}.}
  \bibinfo{year}{2018}\natexlab{}.
\newblock \showarticletitle{Multi-Task Learning for Document Ranking and Query
  Suggestion}. In \bibinfo{booktitle}{{\em ICLR}}.
\newblock


\bibitem[\protect\citeauthoryear{Antonellis, Molina, and Chang}{Antonellis
  et~al\mbox{.}}{2008}]%
        {conf/Simrank}
\bibfield{author}{\bibinfo{person}{Ioannis Antonellis},
  \bibinfo{person}{Hector~Garcia Molina}, {and} \bibinfo{person}{Chi~Chao
  Chang}.} \bibinfo{year}{2008}\natexlab{}.
\newblock \showarticletitle{Simrank++: Query Rewriting Through Link Analysis of
  the Click Graph}. In \bibinfo{booktitle}{{\em VLDB}},
  Vol.~\bibinfo{volume}{1}. \bibinfo{publisher}{VLDB Endowment},
  \bibinfo{pages}{408--421}.
\newblock
\showISSN{2150-8097}


\bibitem[\protect\citeauthoryear{Bai, Zhou, Xue, Zha, Sun, Tseng, Zheng, and
  Chang}{Bai et~al\mbox{.}}{2009}]%
        {Bai-MTL}
\bibfield{author}{\bibinfo{person}{Jing Bai}, \bibinfo{person}{Ke Zhou},
  \bibinfo{person}{Guirong Xue}, \bibinfo{person}{Hongyuan Zha},
  \bibinfo{person}{Gordon Sun}, \bibinfo{person}{Belle Tseng},
  \bibinfo{person}{Zhaohui Zheng}, {and} \bibinfo{person}{Yi Chang}.}
  \bibinfo{year}{2009}\natexlab{}.
\newblock \showarticletitle{Multi-task Learning for Learning to Rank in Web
  Search}. In \bibinfo{booktitle}{{\em Proceedings of the 18th ACM Conference
  on Information and Knowledge Management}} {\em (\bibinfo{series}{CIKM '09})}.
  \bibinfo{publisher}{ACM}, \bibinfo{address}{New York, NY, USA},
  \bibinfo{pages}{1549--1552}.
\newblock
\showISBNx{978-1-60558-512-3}


\bibitem[\protect\citeauthoryear{Boldi, Bonchi, Castillo, Donato, Gionis, and
  Vigna}{Boldi et~al\mbox{.}}{2008}]%
        {conf/query-flow-graph}
\bibfield{author}{\bibinfo{person}{Paolo Boldi}, \bibinfo{person}{Francesco
  Bonchi}, \bibinfo{person}{Carlos Castillo}, \bibinfo{person}{Debora Donato},
  \bibinfo{person}{Aristides Gionis}, {and} \bibinfo{person}{Sebastiano
  Vigna}.} \bibinfo{year}{2008}\natexlab{}.
\newblock \showarticletitle{The query-flow graph: model and applications}. In
  \bibinfo{booktitle}{{\em CIKM '08: Proceeding of the 17th ACM conference on
  Information and knowledge management}}. \bibinfo{publisher}{ACM},
  \bibinfo{address}{New York, NY, USA}, \bibinfo{pages}{609--618}.
\newblock
\showISBNx{978-1-59593-991-3}


\bibitem[\protect\citeauthoryear{Broder, Ciccolo, Gabrilovich, Josifovski,
  Metzler, Riedel, and Yuan}{Broder et~al\mbox{.}}{2009}]%
        {conf/rarequeries}
\bibfield{author}{\bibinfo{person}{Andrei Broder}, \bibinfo{person}{Peter
  Ciccolo}, \bibinfo{person}{Evgeniy Gabrilovich}, \bibinfo{person}{Vanja
  Josifovski}, \bibinfo{person}{Donald Metzler}, \bibinfo{person}{Lance
  Riedel}, {and} \bibinfo{person}{Jeffrey Yuan}.}
  \bibinfo{year}{2009}\natexlab{}.
\newblock \showarticletitle{Online expansion of rare queries for sponsored
  search}. In \bibinfo{booktitle}{{\em WWW '09: Proceedings of the 18th
  international conference on World wide web}}. \bibinfo{publisher}{ACM},
  \bibinfo{address}{New York, NY, USA}, \bibinfo{pages}{511--520}.
\newblock
\showISBNx{978-1-60558-487-4}


\bibitem[\protect\citeauthoryear{Broder, Ciccolo, Fontoura, Gabrilovich,
  Josifovski, and Riedel}{Broder et~al\mbox{.}}{2008}]%
        {conf/cikm/BroderCFGJR08}
\bibfield{author}{\bibinfo{person}{Andrei~Z. Broder}, \bibinfo{person}{Peter
  Ciccolo}, \bibinfo{person}{Marcus Fontoura}, \bibinfo{person}{Evgeniy
  Gabrilovich}, \bibinfo{person}{Vanja Josifovski}, {and}
  \bibinfo{person}{Lance Riedel}.} \bibinfo{year}{2008}\natexlab{}.
\newblock \showarticletitle{Search advertising using web relevance feedback}.
  In \bibinfo{booktitle}{{\em CIKM}} (2008-11-10),
  \bibfield{editor}{\bibinfo{person}{James~G. Shanahan}, \bibinfo{person}{Sihem
  Amer-Yahia}, \bibinfo{person}{Ioana Manolescu}, \bibinfo{person}{Yi~Zhang},
  \bibinfo{person}{David~A. Evans}, \bibinfo{person}{Aleksander Kolcz},
  \bibinfo{person}{Key-Sun Choi}, {and} \bibinfo{person}{Abdur Chowdhury}}
  (Eds.). \bibinfo{publisher}{ACM}, \bibinfo{pages}{1013--1022}.
\newblock
\showISBNx{978-1-59593-991-3}


\bibitem[\protect\citeauthoryear{Cheng and Cant\'u-Paz}{Cheng and
  Cant\'u-Paz}{2010}]%
        {conf/wsdm/ChengC10}
\bibfield{author}{\bibinfo{person}{Haibin Cheng} {and} \bibinfo{person}{Erick
  Cant\'u-Paz}.} \bibinfo{year}{2010}\natexlab{}.
\newblock \showarticletitle{Personalized click prediction in sponsored search}.
  In \bibinfo{booktitle}{{\em WSDM}},
  \bibfield{editor}{\bibinfo{person}{Brian~D. Davison},
  \bibinfo{person}{Torsten Suel}, \bibinfo{person}{Nick Craswell}, {and}
  \bibinfo{person}{Bing Liu}} (Eds.). \bibinfo{publisher}{ACM},
  \bibinfo{pages}{351--360}.
\newblock
\showISBNx{978-1-60558-889-6}


\bibitem[\protect\citeauthoryear{Cheng, Koc, Harmsen, Shaked, Chandra, Aradhye,
  Anderson, Corrado, Chai, Ispir, Anil, Haque, Hong, Jain, Liu, and Shah}{Cheng
  et~al\mbox{.}}{2016}]%
        {conf/widedeep}
\bibfield{author}{\bibinfo{person}{Heng-Tze Cheng}, \bibinfo{person}{Levent
  Koc}, \bibinfo{person}{Jeremiah Harmsen}, \bibinfo{person}{Tal Shaked},
  \bibinfo{person}{Tushar Chandra}, \bibinfo{person}{Hrishi Aradhye},
  \bibinfo{person}{Glen Anderson}, \bibinfo{person}{Greg Corrado},
  \bibinfo{person}{Wei Chai}, \bibinfo{person}{Mustafa Ispir},
  \bibinfo{person}{Rohan Anil}, \bibinfo{person}{Zakaria Haque},
  \bibinfo{person}{Lichan Hong}, \bibinfo{person}{Vihan Jain},
  \bibinfo{person}{Xiaobing Liu}, {and} \bibinfo{person}{Hemal Shah}.}
  \bibinfo{year}{2016}\natexlab{}.
\newblock \showarticletitle{Wide \& Deep Learning for Recommender Systems}. In
  \bibinfo{booktitle}{{\em Proceedings of the 1st Workshop on Deep Learning for
  Recommender Systems}} {\em (\bibinfo{series}{DLRS 2016})}.
  \bibinfo{publisher}{ACM}, \bibinfo{address}{New York, NY, USA},
  \bibinfo{pages}{7--10}.
\newblock
\showISBNx{978-1-4503-4795-2}


\bibitem[\protect\citeauthoryear{Choi, Fontoura, Gabrilovich, Josifovski,
  Mediano, and Pang}{Choi et~al\mbox{.}}{2010}]%
        {conf/www/landingpage}
\bibfield{author}{\bibinfo{person}{Yejin Choi}, \bibinfo{person}{Marcus
  Fontoura}, \bibinfo{person}{Evgeniy Gabrilovich}, \bibinfo{person}{Vanja
  Josifovski}, \bibinfo{person}{Maurício~R. Mediano}, {and}
  \bibinfo{person}{Bo Pang}.} \bibinfo{year}{2010}\natexlab{}.
\newblock \showarticletitle{Using landing pages for sponsored search ad
  selection.}. In \bibinfo{booktitle}{{\em WWW}} (2010-04-28),
  \bibfield{editor}{\bibinfo{person}{Michael Rappa}, \bibinfo{person}{Paul
  Jones}, \bibinfo{person}{Juliana Freire}, {and} \bibinfo{person}{Soumen
  Chakrabarti}} (Eds.). \bibinfo{pages}{251--260}.
\newblock
\showISBNx{978-1-60558-799-8}


\bibitem[\protect\citeauthoryear{Chung, Gulcehre, Cho, and Bengio}{Chung
  et~al\mbox{.}}{2014}]%
        {GRU2ChungGCB14}
\bibfield{author}{\bibinfo{person}{Junyoung Chung}, \bibinfo{person}{Caglar
  Gulcehre}, \bibinfo{person}{Kyunghyun Cho}, {and} \bibinfo{person}{Yoshua
  Bengio}.} \bibinfo{year}{2014}\natexlab{}.
\newblock \bibinfo{booktitle}{{\em Empirical evaluation of gated recurrent
  neural networks on sequence modeling}}.
\newblock


\bibitem[\protect\citeauthoryear{Clevert, Unterthiner, and Hochreiter}{Clevert
  et~al\mbox{.}}{2015}]%
        {elu}
\bibfield{author}{\bibinfo{person}{Djork{-}Arn{\'{e}} Clevert},
  \bibinfo{person}{Thomas Unterthiner}, {and} \bibinfo{person}{Sepp
  Hochreiter}.} \bibinfo{year}{2015}\natexlab{}.
\newblock \showarticletitle{Fast and Accurate Deep Network Learning by
  Exponential Linear Units (ELUs)}.
\newblock \bibinfo{journal}{{\em CoRR\/}}  \bibinfo{volume}{abs/1511.07289}
  (\bibinfo{year}{2015}).
\newblock
\showeprint[arxiv]{1511.07289}


\bibitem[\protect\citeauthoryear{Covington, Adams, and Sargin}{Covington
  et~al\mbox{.}}{2016}]%
        {conf/recsys/YoutubeCovingtonAS16}
\bibfield{author}{\bibinfo{person}{Paul Covington}, \bibinfo{person}{Jay
  Adams}, {and} \bibinfo{person}{Emre Sargin}.}
  \bibinfo{year}{2016}\natexlab{}.
\newblock \showarticletitle{Deep Neural Networks for YouTube Recommendations}.
  In \bibinfo{booktitle}{{\em RecSys}},
  \bibfield{editor}{\bibinfo{person}{Shilad Sen}, \bibinfo{person}{Werner
  Geyer}, \bibinfo{person}{Jill Freyne}, {and} \bibinfo{person}{Pablo
  Castells}} (Eds.). \bibinfo{publisher}{ACM}, \bibinfo{pages}{191--198}.
\newblock
\showISBNx{978-1-4503-4035-9}


\bibitem[\protect\citeauthoryear{Duchi, Hazan, and Singer}{Duchi
  et~al\mbox{.}}{2011}]%
        {adagrad}
\bibfield{author}{\bibinfo{person}{John Duchi}, \bibinfo{person}{Elad Hazan},
  {and} \bibinfo{person}{Yoram Singer}.} \bibinfo{year}{2011}\natexlab{}.
\newblock \showarticletitle{Adaptive Subgradient Methods for Online Learning
  and Stochastic Optimization}.
\newblock \bibinfo{journal}{{\em J. Mach. Learn. Res.\/}}  \bibinfo{volume}{12}
  (\bibinfo{date}{July} \bibinfo{year}{2011}), \bibinfo{pages}{2121--2159}.
\newblock
\showISSN{1532-4435}


\bibitem[\protect\citeauthoryear{Grbovic, Djuric, Radosavljevic, Silvestri,
  Baeza-Yates, Feng, Ordentlich, Yang, and Owens}{Grbovic
  et~al\mbox{.}}{2016}]%
        {grbovic2016scalable}
\bibfield{author}{\bibinfo{person}{Mihajlo Grbovic}, \bibinfo{person}{Nemanja
  Djuric}, \bibinfo{person}{Vladan Radosavljevic}, \bibinfo{person}{Fabrizio
  Silvestri}, \bibinfo{person}{Ricardo Baeza-Yates}, \bibinfo{person}{Andrew
  Feng}, \bibinfo{person}{Erik Ordentlich}, \bibinfo{person}{Lee Yang}, {and}
  \bibinfo{person}{Gavin Owens}.} \bibinfo{year}{2016}\natexlab{}.
\newblock \showarticletitle{Scalable Semantic Matching of Queries to Ads in
  Sponsored Search Advertising}. In \bibinfo{booktitle}{{\em SIGIR}}.
  \bibinfo{publisher}{ACM}, \bibinfo{address}{New York, NY, USA},
  \bibinfo{pages}{375--384}.
\newblock
\showISBNx{978-1-4503-4069-4}


\bibitem[\protect\citeauthoryear{Grbovic, Djuric, Radosavljevic, Silvestri, and
  Bhamidipati}{Grbovic et~al\mbox{.}}{2015}]%
        {conf/sigir/GrbovicDRSB15_queryrewrite}
\bibfield{author}{\bibinfo{person}{Mihajlo Grbovic}, \bibinfo{person}{Nemanja
  Djuric}, \bibinfo{person}{Vladan Radosavljevic}, \bibinfo{person}{Fabrizio
  Silvestri}, {and} \bibinfo{person}{Narayan Bhamidipati}.}
  \bibinfo{year}{2015}\natexlab{}.
\newblock \showarticletitle{Context- and Content-aware Embeddings for Query
  Rewriting in Sponsored Search}. In \bibinfo{booktitle}{{\em SIGIR}}.
  \bibinfo{publisher}{ACM}, \bibinfo{pages}{383--392}.
\newblock
\showISBNx{978-1-4503-3621-5}


\bibitem[\protect\citeauthoryear{Guo, Fan, Ai, and Croft}{Guo
  et~al\mbox{.}}{2016}]%
        {Guo:2016}
\bibfield{author}{\bibinfo{person}{Jiafeng Guo}, \bibinfo{person}{Yixing Fan},
  \bibinfo{person}{Qingyao Ai}, {and} \bibinfo{person}{W.~Bruce Croft}.}
  \bibinfo{year}{2016}\natexlab{}.
\newblock \showarticletitle{Semantic Matching by Non-Linear Word Transportation
  for Information Retrieval}. In \bibinfo{booktitle}{{\em Proceedings of the
  25th ACM International on Conference on Information and Knowledge
  Management}} {\em (\bibinfo{series}{CIKM '16})}. \bibinfo{publisher}{ACM},
  \bibinfo{address}{New York, NY, USA}, \bibinfo{pages}{701--710}.
\newblock
\showISBNx{978-1-4503-4073-1}


\bibitem[\protect\citeauthoryear{He, Zhang, Ren, and Sun}{He
  et~al\mbox{.}}{2015}]%
        {prelu}
\bibfield{author}{\bibinfo{person}{Kaiming He}, \bibinfo{person}{Xiangyu
  Zhang}, \bibinfo{person}{Shaoqing Ren}, {and} \bibinfo{person}{Jian Sun}.}
  \bibinfo{year}{2015}\natexlab{}.
\newblock \showarticletitle{Delving Deep into Rectifiers: Surpassing
  Human-Level Performance on ImageNet Classification}.
\newblock \bibinfo{journal}{{\em CoRR\/}}  \bibinfo{volume}{abs/1502.01852}
  (\bibinfo{year}{2015}).
\newblock
\showeprint{1502.01852}


\bibitem[\protect\citeauthoryear{Huang, He, Gao, Deng, Acero, and Heck}{Huang
  et~al\mbox{.}}{2013}]%
        {conf/cikm/HuangHGDAH13_dssm}
\bibfield{author}{\bibinfo{person}{Po-Sen Huang}, \bibinfo{person}{Xiaodong
  He}, \bibinfo{person}{Jianfeng Gao}, \bibinfo{person}{Li Deng},
  \bibinfo{person}{Alex Acero}, {and} \bibinfo{person}{Larry~P. Heck}.}
  \bibinfo{year}{2013}\natexlab{}.
\newblock \showarticletitle{Learning deep structured semantic models for web
  search using clickthrough data}. In \bibinfo{booktitle}{{\em CIKM}}.
  \bibinfo{publisher}{ACM}, \bibinfo{pages}{2333--2338}.
\newblock
\showISBNx{978-1-4503-2263-8}


\bibitem[\protect\citeauthoryear{Johnson, Douze, and J{\'e}gou}{Johnson
  et~al\mbox{.}}{2017}]%
        {JDH17}
\bibfield{author}{\bibinfo{person}{Jeff Johnson}, \bibinfo{person}{Matthijs
  Douze}, {and} \bibinfo{person}{Herv{\'e} J{\'e}gou}.}
  \bibinfo{year}{2017}\natexlab{}.
\newblock \showarticletitle{Billion-scale similarity search with GPUs}.
\newblock \bibinfo{journal}{{\em arXiv preprint arXiv:1702.08734\/}}
  (\bibinfo{year}{2017}).
\newblock


\bibitem[\protect\citeauthoryear{Klambauer, Unterthiner, Mayr, and
  Hochreiter}{Klambauer et~al\mbox{.}}{2017}]%
        {selu}
\bibfield{author}{\bibinfo{person}{Günter Klambauer}, \bibinfo{person}{Thomas
  Unterthiner}, \bibinfo{person}{Andreas Mayr}, {and} \bibinfo{person}{Sepp
  Hochreiter}.} \bibinfo{year}{2017}\natexlab{}.
\newblock \showarticletitle{Self-Normalizing Neural Networks}.
\newblock \bibinfo{journal}{{\em CoRR\/}}  \bibinfo{volume}{abs/1706.02515}
  (\bibinfo{year}{2017}).
\newblock


\bibitem[\protect\citeauthoryear{Liu, Xiao, Ou, and Si}{Liu
  et~al\mbox{.}}{2017}]%
        {Liu:2017:CRO:3097983.3098011}
\bibfield{author}{\bibinfo{person}{Shichen Liu}, \bibinfo{person}{Fei Xiao},
  \bibinfo{person}{Wenwu Ou}, {and} \bibinfo{person}{Luo Si}.}
  \bibinfo{year}{2017}\natexlab{}.
\newblock \showarticletitle{Cascade Ranking for Operational E-commerce Search}.
  In \bibinfo{booktitle}{{\em Proceedings of the 23rd ACM SIGKDD International
  Conference on Knowledge Discovery and Data Mining}} {\em
  (\bibinfo{series}{KDD '17})}. \bibinfo{publisher}{ACM}, \bibinfo{address}{New
  York, NY, USA}, \bibinfo{pages}{1557--1565}.
\newblock
\showISBNx{978-1-4503-4887-4}


\bibitem[\protect\citeauthoryear{Liu, Gao, He, Deng, Duh, and Wang}{Liu
  et~al\mbox{.}}{2015}]%
        {conf/naacl/LiuGHDDW15}
\bibfield{author}{\bibinfo{person}{Xiaodong Liu}, \bibinfo{person}{Jianfeng
  Gao}, \bibinfo{person}{Xiaodong He}, \bibinfo{person}{Li Deng},
  \bibinfo{person}{Kevin Duh}, {and} \bibinfo{person}{Ye{-}Yi Wang}.}
  \bibinfo{year}{2015}\natexlab{}.
\newblock \showarticletitle{Representation Learning Using Multi-Task Deep
  Neural Networks for Semantic Classification and Information Retrieval}. In
  \bibinfo{booktitle}{{\em {NAACL} {HLT} 2015, The 2015 Conference of the North
  American Chapter of the Association for Computational Linguistics: Human
  Language Technologies, Denver, Colorado, USA, May 31 - June 5, 2015}}.
  \bibinfo{pages}{912--921}.
\newblock


\bibitem[\protect\citeauthoryear{Malekian, Chang, Kumar, and Wang}{Malekian
  et~al\mbox{.}}{2008}]%
        {nb}
\bibfield{author}{\bibinfo{person}{Azarakhsh Malekian},
  \bibinfo{person}{Chi-Chao Chang}, \bibinfo{person}{Ravi Kumar}, {and}
  \bibinfo{person}{Grant Wang}.} \bibinfo{year}{2008}\natexlab{}.
\newblock \showarticletitle{Optimizing Query Rewrites for Keyword-based
  Advertising}. In \bibinfo{booktitle}{{\em Proceedings of the 9th ACM
  Conference on Electronic Commerce}} {\em (\bibinfo{series}{EC '08})}.
  \bibinfo{publisher}{ACM}, \bibinfo{address}{New York, NY, USA},
  \bibinfo{pages}{10--19}.
\newblock
\showISBNx{978-1-60558-169-9}


\bibitem[\protect\citeauthoryear{Mikolov, Sutskever, Chen, Corrado, and
  Dean}{Mikolov et~al\mbox{.}}{2013}]%
        {conf/mikolov2013distributed}
\bibfield{author}{\bibinfo{person}{Tomas Mikolov}, \bibinfo{person}{Ilya
  Sutskever}, \bibinfo{person}{Kai Chen}, \bibinfo{person}{Greg~S Corrado},
  {and} \bibinfo{person}{Jeff Dean}.} \bibinfo{year}{2013}\natexlab{}.
\newblock \showarticletitle{Distributed Representations of Words and Phrases
  and their Compositionality}. In \bibinfo{booktitle}{{\em NIPS}}.
  \bibinfo{publisher}{Curran Associates, Inc.}, \bibinfo{pages}{3111--3119}.
\newblock


\bibitem[\protect\citeauthoryear{Mitra, Diaz, and Craswell}{Mitra
  et~al\mbox{.}}{2017}]%
        {Mitra:2017}
\bibfield{author}{\bibinfo{person}{Bhaskar Mitra}, \bibinfo{person}{Fernando
  Diaz}, {and} \bibinfo{person}{Nick Craswell}.}
  \bibinfo{year}{2017}\natexlab{}.
\newblock \showarticletitle{Learning to Match Using Local and Distributed
  Representations of Text for Web Search}. In \bibinfo{booktitle}{{\em
  Proceedings of the 26th International Conference on World Wide Web}} {\em
  (\bibinfo{series}{WWW '17})}. \bibinfo{pages}{1291--1299}.
\newblock


\bibitem[\protect\citeauthoryear{Nair and Hinton}{Nair and Hinton}{2010}]%
        {relu}
\bibfield{author}{\bibinfo{person}{Vinod Nair} {and}
  \bibinfo{person}{Geoffrey~E. Hinton}.} \bibinfo{year}{2010}\natexlab{}.
\newblock \showarticletitle{Rectified Linear Units Improve Restricted Boltzmann
  Machines}. In \bibinfo{booktitle}{{\em ICML}},
  \bibfield{editor}{\bibinfo{person}{Johannes Fürnkranz} {and}
  \bibinfo{person}{Thorsten Joachims}} (Eds.). \bibinfo{publisher}{Omnipress},
  \bibinfo{pages}{807--814}.
\newblock


\bibitem[\protect\citeauthoryear{Okura, Tagami, Ono, and Tajima}{Okura
  et~al\mbox{.}}{2017}]%
        {Okura2017}
\bibfield{author}{\bibinfo{person}{Shumpei Okura}, \bibinfo{person}{Yukihiro
  Tagami}, \bibinfo{person}{Shingo Ono}, {and} \bibinfo{person}{Akira Tajima}.}
  \bibinfo{year}{2017}\natexlab{}.
\newblock \showarticletitle{Embedding-based News Recommendation for Millions of
  Users}. In \bibinfo{booktitle}{{\em Proceedings of the 23rd ACM SIGKDD
  International Conference on Knowledge Discovery and Data Mining}} {\em
  (\bibinfo{series}{KDD '17})}. \bibinfo{publisher}{ACM}, \bibinfo{address}{New
  York, NY, USA}, \bibinfo{pages}{1933--1942}.
\newblock
\showISBNx{978-1-4503-4887-4}


\bibitem[\protect\citeauthoryear{Palangi, Deng, Shen, Gao, He, Chen, Song, and
  Ward}{Palangi et~al\mbox{.}}{2015}]%
        {journals/rnn-dssm}
\bibfield{author}{\bibinfo{person}{Hamid Palangi}, \bibinfo{person}{Li Deng},
  \bibinfo{person}{Yelong Shen}, \bibinfo{person}{Jianfeng Gao},
  \bibinfo{person}{Xiaodong He}, \bibinfo{person}{Jianshu Chen},
  \bibinfo{person}{Xinying Song}, {and} \bibinfo{person}{Rabab~K. Ward}.}
  \bibinfo{year}{2015}\natexlab{}.
\newblock \showarticletitle{Deep Sentence Embedding Using the Long Short Term
  Memory Network: Analysis and Application to Information Retrieval.}
\newblock \bibinfo{journal}{{\em CoRR\/}}  \bibinfo{volume}{abs/1502.06922}
  (\bibinfo{year}{2015}).
\newblock


\bibitem[\protect\citeauthoryear{Pang, Lan, Guo, Xu, Wan, and Cheng}{Pang
  et~al\mbox{.}}{2016}]%
        {Pang:2016}
\bibfield{author}{\bibinfo{person}{Liang Pang}, \bibinfo{person}{Yanyan Lan},
  \bibinfo{person}{Jiafeng Guo}, \bibinfo{person}{Jun Xu},
  \bibinfo{person}{Shengxian Wan}, {and} \bibinfo{person}{Xueqi Cheng}.}
  \bibinfo{year}{2016}\natexlab{}.
\newblock \showarticletitle{Text Matching As Image Recognition}. In
  \bibinfo{booktitle}{{\em Proceedings of the Thirtieth AAAI Conference on
  Artificial Intelligence}} {\em (\bibinfo{series}{AAAI'16})}.
  \bibinfo{publisher}{AAAI Press}, \bibinfo{pages}{2793--2799}.
\newblock


\bibitem[\protect\citeauthoryear{Ramachandran, Zoph, and Le}{Ramachandran
  et~al\mbox{.}}{2017}]%
        {swish}
\bibfield{author}{\bibinfo{person}{Prajit Ramachandran},
  \bibinfo{person}{Barret Zoph}, {and} \bibinfo{person}{Quoc~V. Le}.}
  \bibinfo{year}{2017}\natexlab{}.
\newblock \showarticletitle{Searching for Activation Functions}.
\newblock \bibinfo{journal}{{\em CoRR\/}}  \bibinfo{volume}{abs/1710.05941}
  (\bibinfo{year}{2017}).
\newblock
\showeprint[arxiv]{1710.05941}


\bibitem[\protect\citeauthoryear{Ruder}{Ruder}{2017}]%
        {journals/corr/Ruder17a}
\bibfield{author}{\bibinfo{person}{Sebastian Ruder}.}
  \bibinfo{year}{2017}\natexlab{}.
\newblock \showarticletitle{An Overview of Multi-Task Learning in Deep Neural
  Networks}.
\newblock \bibinfo{journal}{{\em CoRR\/}}  \bibinfo{volume}{abs/1706.05098}
  (\bibinfo{year}{2017}).
\newblock


\bibitem[\protect\citeauthoryear{Shan, Hoens, Jiao, Wang, Yu, and Mao}{Shan
  et~al\mbox{.}}{2016}]%
        {conf/deepcrossing}
\bibfield{author}{\bibinfo{person}{Ying Shan}, \bibinfo{person}{T.~Ryan Hoens},
  \bibinfo{person}{Jian Jiao}, \bibinfo{person}{Haijing Wang},
  \bibinfo{person}{Dong Yu}, {and} \bibinfo{person}{JC Mao}.}
  \bibinfo{year}{2016}\natexlab{}.
\newblock \showarticletitle{Deep Crossing: Web-Scale Modeling Without Manually
  Crafted Combinatorial Features}. In \bibinfo{booktitle}{{\em Proceedings of
  the 22Nd ACM SIGKDD International Conference on Knowledge Discovery and Data
  Mining}} {\em (\bibinfo{series}{KDD '16})}. \bibinfo{publisher}{ACM},
  \bibinfo{address}{New York, NY, USA}, \bibinfo{pages}{255--262}.
\newblock
\showISBNx{978-1-4503-4232-2}


\bibitem[\protect\citeauthoryear{Shen, He, Gao, Deng, and Mesnil}{Shen
  et~al\mbox{.}}{2014}]%
        {conf/cdssm}
\bibfield{author}{\bibinfo{person}{Yelong Shen}, \bibinfo{person}{Xiaodong He},
  \bibinfo{person}{Jianfeng Gao}, \bibinfo{person}{Li Deng}, {and}
  \bibinfo{person}{Grégoire Mesnil}.} \bibinfo{year}{2014}\natexlab{}.
\newblock \showarticletitle{Learning semantic representations using
  convolutional neural networks for web search}. In \bibinfo{booktitle}{{\em
  WWW (Companion Volume)}}, \bibfield{editor}{\bibinfo{person}{Chin-Wan Chung},
  \bibinfo{person}{Andrei~Z. Broder}, \bibinfo{person}{Kyuseok Shim}, {and}
  \bibinfo{person}{Torsten Suel}} (Eds.). \bibinfo{publisher}{ACM},
  \bibinfo{pages}{373--374}.
\newblock


\bibitem[\protect\citeauthoryear{Su, Wei, Tianshu, Daorui, Bo, and Kaipeng}{Su
  et~al\mbox{.}}{2018}]%
        {conf/www2017/Yansu}
\bibfield{author}{\bibinfo{person}{Yan Su}, \bibinfo{person}{Lin Wei},
  \bibinfo{person}{Wu Tianshu}, \bibinfo{person}{Xiao Daorui},
  \bibinfo{person}{Wu Bo}, {and} \bibinfo{person}{Liu Kaipeng}.}
  \bibinfo{year}{2018}\natexlab{}.
\newblock \showarticletitle{Beyond Keywords and Relevance: A Personalized Ad
  Retrieval Framework in E-Commerce Sponsored Search}.
\newblock  (\bibinfo{year}{2018}).
\newblock
\showURL{%
\url{https://arxiv.org/abs/1712.10110}}


\bibitem[\protect\citeauthoryear{Sukhbaatar, Szlam, Weston, and
  Fergus}{Sukhbaatar et~al\mbox{.}}{2015}]%
        {memorynetwork}
\bibfield{author}{\bibinfo{person}{Sainbayar Sukhbaatar},
  \bibinfo{person}{Arthur Szlam}, \bibinfo{person}{Jason Weston}, {and}
  \bibinfo{person}{Rob Fergus}.} \bibinfo{year}{2015}\natexlab{}.
\newblock \showarticletitle{End-To-End Memory Networks}. In
  \bibinfo{booktitle}{{\em NIPS}}. \bibinfo{pages}{2440--2448}.
\newblock


\bibitem[\protect\citeauthoryear{Sussillo}{Sussillo}{2014}]%
        {journals/corr/Sussillo14}
\bibfield{author}{\bibinfo{person}{David Sussillo}.}
  \bibinfo{year}{2014}\natexlab{}.
\newblock \showarticletitle{Random Walks: Training Very Deep Nonlinear
  Feed-Forward Networks with Smart Initialization}.
\newblock \bibinfo{journal}{{\em CoRR\/}}  \bibinfo{volume}{abs/1412.6558}
  (\bibinfo{year}{2014}).
\newblock


\bibitem[\protect\citeauthoryear{Zhai, Chang, Zhang, and Zhang}{Zhai
  et~al\mbox{.}}{2016}]%
        {Zhai:2016:DLA:2939672.2939759}
\bibfield{author}{\bibinfo{person}{Shuangfei Zhai}, \bibinfo{person}{Keng-hao
  Chang}, \bibinfo{person}{Ruofei Zhang}, {and} \bibinfo{person}{Zhongfei~Mark
  Zhang}.} \bibinfo{year}{2016}\natexlab{}.
\newblock \showarticletitle{DeepIntent: Learning Attentions for Online
  Advertising with Recurrent Neural Networks}. In \bibinfo{booktitle}{{\em
  Proceedings of the 22Nd ACM SIGKDD International Conference on Knowledge
  Discovery and Data Mining}} {\em (\bibinfo{series}{KDD '16})}.
  \bibinfo{publisher}{ACM}, \bibinfo{address}{New York, NY, USA},
  \bibinfo{pages}{1295--1304}.
\newblock
\showISBNx{978-1-4503-4232-2}


\bibitem[\protect\citeauthoryear{Zhang, Dai, Xu, Feng, Wang, Bian, Wang, and
  Liu}{Zhang et~al\mbox{.}}{2014}]%
        {conf/rnnctr}
\bibfield{author}{\bibinfo{person}{Yuyu Zhang}, \bibinfo{person}{Hanjun Dai},
  \bibinfo{person}{Chang Xu}, \bibinfo{person}{Jun Feng},
  \bibinfo{person}{Taifeng Wang}, \bibinfo{person}{Jiang Bian},
  \bibinfo{person}{Bin Wang}, {and} \bibinfo{person}{Tie-Yan Liu}.}
  \bibinfo{year}{2014}\natexlab{}.
\newblock \showarticletitle{Sequential Click Prediction for Sponsored Search
  with Recurrent Neural Networks}. In \bibinfo{booktitle}{{\em AAAI}}.
  \bibinfo{publisher}{AAAI Press}, \bibinfo{pages}{1369--1375}.
\newblock
\showISBNx{978-1-57735-661-5}


\bibitem[\protect\citeauthoryear{Zhou, Song, Zhu, Ma, Yan, Dai, Zhu, Jin, Li,
  and Gai}{Zhou et~al\mbox{.}}{2017}]%
        {deepinterest}
\bibfield{author}{\bibinfo{person}{Guorui Zhou}, \bibinfo{person}{Chengru
  Song}, \bibinfo{person}{Xiaoqiang Zhu}, \bibinfo{person}{Xiao Ma},
  \bibinfo{person}{Yanghui Yan}, \bibinfo{person}{Xingya Dai},
  \bibinfo{person}{Han Zhu}, \bibinfo{person}{Junqi Jin}, \bibinfo{person}{Han
  Li}, {and} \bibinfo{person}{Kun Gai}.} \bibinfo{year}{2017}\natexlab{}.
\newblock \showarticletitle{Deep Interest Network for Click-Through Rate
  Prediction}.
\newblock  (\bibinfo{year}{2017}).
\newblock
\showURL{%
\url{https://arxiv.org/abs/1706.06978}}


\end{thebibliography}

\end{document}